\documentclass[onecolumn
]{ceurart}

\usepackage{listings}
\usepackage{float}
\usepackage{graphicx}
\usepackage{subcaption}
\begin{document}

\copyrightyear{2024}
\copyrightclause{Copyright for this paper by its authors.
  Use permitted under Creative Commons License Attribution 4.0
  International (CC BY 4.0).}

\conference{LAK'24: International Workshop on Generative AI for Learning Analytics (GenAI-LA),
  March 19, 2024, Kyoto, Japan}

\title{3DG: A Framework for Using Generative AI for Handling Sparse Learner Performance Data From Intelligent Tutoring Systems}

\tnotemark[1]

\author[1,2]{Liang Zhang}[%
orcid=0009-0002-0017-2569,
email=lzhang13@memphis.edu]
\address[1]{Institute for Intelligent Systems, University of Memphis, Memphis, TN 38152, USA}
\address[2]{Department of Electrical and Computer Engineering, University of Memphis, Memphis, TN 38152, USA}

\author[3]{Jionghao Lin}[%
orcid=0000-0003-3320-3907,
email=jionghao@cmu.edu]
\cormark[1]
\address[3]{Human-Computer Interaction Institute, Carnegie Mellon University, Pittsburgh, PA, 15213, USA}

\author[3]{Conrad Borchers}[%
orcid=0000-0003-3437-8979,
email=cborcher@cs.cmu.edu]

\author[4]{Meng Cao}[%
orcid=0000-0002-1286-2885,
email=mcao@memphis.edu]
\address[4]{Department of Psychology, University of Memphis, Memphis, TN, 38152, USA}

\author[1,2,4]{Xiangen Hu}[%
orcid=0000-0001-9045-4070,
email=xhu@memphis.edu]

\cortext[1]{Corresponding author.}

\begin{abstract}
Learning performance data (e.g., quiz scores and attempts) is significant for understanding learner engagement and knowledge mastery level. However, the learning performance data collected from Intelligent Tutoring Systems (ITSs) often suffers from sparsity, impacting the accuracy of learner modeling and knowledge assessments. To address this, we introduce the 3DG framework (3-Dimensional tensor for Densification and Generation), a novel approach combining tensor factorization with advanced generative models, including Generative Adversarial Network (GAN) and Generative Pre-trained Transformer (GPT), for enhanced data imputation and augmentation. The framework operates by first representing the data as a three-dimensional tensor, capturing dimensions of learners, questions, and attempts. It then densifies the data through tensor factorization and augments it using Generative AI models, tailored to individual learning patterns identified via clustering. Applied to data from an AutoTutor lesson by the Center for the Study of Adult Literacy (CSAL), the 3DG framework effectively generated scalable, personalized simulations of learning performance. Comparative analysis revealed GAN's superior reliability over GPT-4 in this context, underscoring its potential in addressing data sparsity challenges in ITSs and contributing to the advancement of personalized educational technology.
\end{abstract}


\maketitle

\section{Introduction}

Intelligent Tutoring System (ITS) is a prototype of computer system designed to offer personalized and adaptive instructions through tracing and analyzing learning performance data such as quiz scores and question attempts \cite{anderson1985intelligent,corbett1997intelligent,vanlehn2011relative,graesser2018intelligent}. However, during the interaction between learners and ITS, learning performance data often exhibits data sparsity due to unexplored questions, insufficient attempts to master knowledge, and lacking variability in learning patterns \cite{nguyen2011predicting,pandey2019self,zhang2020attention,wang2021knowledge,wang2023graphca}. Data sparsity can lead to biased analysis and modeling of learning data. This is particularly evident in the ``Learner Model'' component of ITS, which is crucial for tracking learning and predicting performance of individual learners \cite{pavlik2021automatic,eglington2023optimize,zhang2023exploring}. Specifically, sparse performance data can lead to skewed or overfitted Knowledge Tracing models in ``Learner Model'', which impedes accurately capturing learner knowledge states and may result in misleading predictions of learning performance \cite{pandey2019self,wang2023graphca,wang2019deep,lee2022contrastive}. The scarcity of learning performance data significantly hampers the development of ITSs, particularly in cases where learners have not sufficiently engaged with certain instructional scenarios \cite{kossiakoff2011systems,baudin2017openturns}.

Tackling data sparsity for ITSs presents a practical yet challenging research area. Informed by the machine learning literature \cite{acar2011scalable,shorten2019survey,emmanuel2021survey,liu2021adaptive}, the issue of data sparsity can be addressed by two principal ways: \textit{data imputation} and \textit{data augmentation}. Firstly, data imputation focuses on filling the gaps in missing data to ensure a comprehensive dataset \cite{pandey2019self,wang2021knowledge,thai2012factorization}. Secondly, data augmentation aims to enrich and expand datasets where there are insufficient learning patterns, thus ensuring robustness in analysis, modeling, and even potential testing tasks for ITSs \cite{wang2023graphca,krizhevsky2012imagenet}. Currently, limited efforts have been made in the field of ITSs to systematically address these data sparsity issues in learning performance data \cite{nguyen2011predicting,pandey2019self,thai2012factorization}. Driven by this, we propose the 3DG (\textbf{3}-dimensions, \textbf{D}ensification, and \textbf{G}eneration) simulation framework, a systematic approach leveraging generative models to handle sparse learning performance from ITS. 

The 3DG framework derived from its three core phases. In the \textit{first phase}, a 3-dimensional tensor is constructed to represent learning performance data, with dimensions corresponding to learners, questions, and attempts. The \textit{second phase} focuses on densifying the sparse tensor by tensor factorization. The \textit{third phase} entails the generation of learning performance data based on generative models, tailored to the individual learning patterns of learners. The 3DG framework integrates the multidimensional learner model with generative models to facilitate scalable simulation sampling for individual learning patterns. The multidimensional learner model in our framework is derived from the Tensor Factorization method, a widely-used approach in predicting learner performance in many studies \cite{thai2012factorization,sahebi2016tensor,doan2019rank,zhao2020modeling}. Initially, learning performance values are represented in a three-dimensional tensor encompassing dimensions of learners, questions, and attempts. Specifically, learning performance indicators, such as binary responses from learners at problem-solving step attempts (with correct answers denoted as 1 and incorrect as 0), form the tensor entries, and they are arranged sequentially along the question queue in the learning process and sorted by attempts in ascending order. This constructed tensor exhibits data sparsity. Our study aims to perform data imputation and augmentation on the sparse tensor. Mathematically, the tensor factorization method addresses incomplete and missing performance values in factorization computations, serving as a form of tensor completion typically used in data imputation  \cite{thai2012factorization,thai2011factorization,chen2019missing}). Inspired by the recent advancements of generative models \cite{jovanovic2022generative,baidoo2023education}, which are capable of generating data based on patterns learned during training and have revolutionized simulation methodologies to be more flexible and cost-effective, our study delves into exploring their potential of addressing the data sparsity issue. We operate under the foundational assumption that, if learning patterns can be identified within the multidimensional learner model, they can be effectively simulated and generated using generative models, facilitating scalable data augmentation. Consequently, current research was guided by following two \textbf{R}esearch \textbf{Q}uestions: 

\begin{itemize}
    \item \textbf{RQ 1:} What is the most effective method for integrating tensor factorization and generative models to develop a systematic framework that proficiently imputes and augments sparse learning performance data?
    \item \textbf{RQ 2:} In the context of simulating learning performance data, how do Generative Adversarial Network (GAN) and Generative Pre-trained Transformer (GPT) models compare in terms of effectiveness and accuracy?
\end{itemize}

\section{Methods}
\subsection{Dataset}

Our study investigated a dataset derived from the AutoTutor ITS, focusing on learning performance in reading comprehension. This dataset originates from lessons developed for the Center for the Study of Adult Literacy (CSAL) \cite{graesser2016reading,fang2018clustering}, specifically the 'Cause and Effect' lesson, involving 118 participants. The lesson design incorporates three levels of question difficulty: medium (M), easy (E), and hard (H). There are 9 medium-difficulty questions, 10 easy questions, and 10 hard questions. Notably, the distribution of learners across these difficulty levels varies within the lesson. Upon completing the medium difficulty level, learners are either advanced to the hard level or redirected to the easy level, depending on their performance, thus providing a tailored learning pathway.

\subsection{The Systematic Simulation Framework}

\begin{figure}[b]
\includegraphics[width=\textwidth]{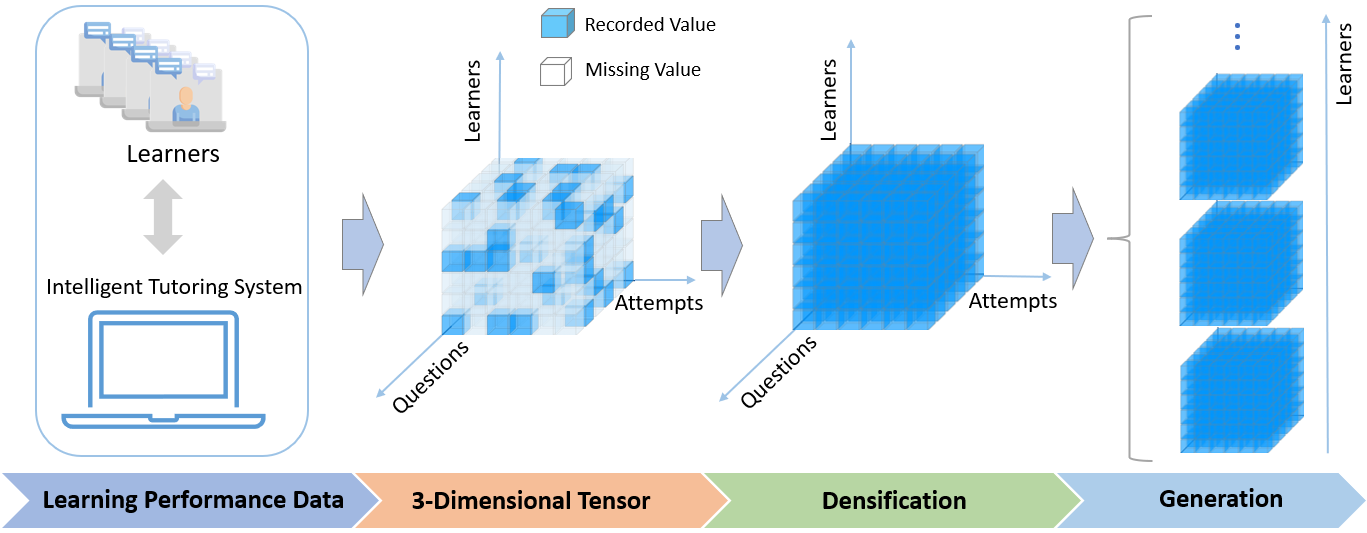}
\caption{The \textbf{3}-dimensions, \textbf{D}ensification, and \textbf{G}eneration (3DG) systematic simulation framework.} \label{framework}
\end{figure}

We propose a systematic simulation framework, 3DG, illustrated in Figure \ref{framework}. This framework begins by structuring the initial learning performance data, sourcing from real-world learner-ITS interactions, into a three-dimensional tensor by dimensions of learners, questions, and attempts. As depicted in the sparse cube space in Figure \ref{framework}, filled cubes represent recorded values of learning performance, while transparent cubes indicate missing values. Tensor completion (based on tensor factorization) is then utilized, converting the sparse tensor to a densified one. The densified tensor provides invaluable information in identifying various learning patterns, which aids in dividing the tensor into sub-tensors by categorizing distinct learning patterns. Subsequently, generative models are harnessed to simulate additional data samples for enriching the original dataset based on each specific learning pattern. The entire operation is encapsulated for scalable simulation sampling and ultimately offers a comprehensive dataset incorporating both imputed and augmented data. This framework was developed to address \textbf{RQ} 1. More detailed methods within this framework are described in the following subsections.

\subsection{Tensor Completion for Data Imputation}

The three-dimensional tensor \(\boldsymbol{\mathcal{T}}\), representing the learning process, is defined as \(\boldsymbol{\mathcal{T}}\in R^{U\times N \times M }\), where the \(U=max(1,2,3,\cdots,u)\) is the maximum number of learners, \(N=max(1,2,3,\cdots,n)\) the maximum number of questions, and \(M=max(1,2,3,\cdots,m)\) the maximum number of attempts. Each element \(\tau_{uij}\) of \(\boldsymbol{\mathcal{T}}\) indicates the performance variable of learner \(l_u\) on question \(q_{i}\) at the attempt \(a_j\). For instance, in the CSAL AutoTutor context, a binary variable \(\boldsymbol{\tau}_{uij}=\{0,1\}\) is used, where 1 signifies a correct answer and 0 denotes an incorrect one. We model the tenor \(\boldsymbol{\mathcal{T}}\) as a factorization of two lower dimensional components: 1) a learner latent matrix \(\boldsymbol{\mathcal{U}}\) of size \textbf{\(U\times K\)} (\(K\) represents the set of latent features in tensor factorization), which captures learner-related latent features matrix/space (such as abilities and learning-related features); and 2) a latent tensor \(\boldsymbol{\mathcal{V}}\) of size \(K\times M \times N\), representing the learner knowledge in terms of latent features during question attempts. The approximated tensor \(\hat{\boldsymbol{\mathcal{T}}}\) is obtained by the following formula:
\begin{equation} 
\hat{\boldsymbol{\mathcal{T}}}\approx \boldsymbol{\mathcal{U}} \times \boldsymbol{\mathcal{V}}
\end{equation}
where \(\boldsymbol{\mathcal{U}}\) can be interpreted as the latent feature space encapsulating learner-related effects, reflecting characteristics such as individual abilities/features and learning preferences. On the other hand, the tensor \(\boldsymbol{\mathcal{V}}\) represents the interaction between attempts and question-related (knowledge acquisition) effects, adapting to various learner features. 

\subsection{Scalable Simulation based on Generative Models for Data Augmentation}

To answer \textbf{RQ} 2, we used two generative models, GAN (Generative Adversarial Network) and GPT (Generative Pre-trained Transformer), to facilitate scalable simulations that are tailored to individual learning patterns. According to \cite{goodfellow2014generative,goodfellow2020generative}, GAN model is uniquely structured with a dual-network architecture comprising a generator and a discriminator. This architecture enables GAN model to excel in generating high-quality synthetic data. In comparison, GPT model is  distinguished by its use of transformer architecture, which empowers it to generate data that is not only contextually relevant but also maintains a high degree of coherence \cite{vaswani2017attention,radford2018improving}.

Before initiating the simulations, we employ a clustering algorithm (i.e., K-means++) to categorize individual learning patterns based on similarities in learners' performance. The learners-attempts matrix slice extracted from the \(\hat{\boldsymbol{\mathcal{T}}}\), encapsulates the probability-based knowledge states associated with the performance on the \(n\)th question \(q_n\), for all \(U\) learners over \(M\) attempts. In our analysis, we employ the \textit{``power law learning curve"}, a model widely recognized in educational and training research \cite{newell1980mechanisms,cen2006learning,dekeyser2020skill}, to fit the learning performance with increasing attempts. In the power-law formula \(Y=aX^{b}\), the \(Y\) represents the learning performance, quantified as the probability of producing correct answers, and \(X\) is the number of opportunities to practice a skill or attempt. The parameter \(a\) indicates the measurement of the learner's initial ability or prior knowledge, and \(b\) represents the learning rate at which the learner acquires knowledge through practice. We employ K-means++ \cite{arthur2007k,bahmani2012scalable} to cluster the distribution of two model parameters (\(a\) and \(b\)), which assists in identifying distinct individual learning patterns.

\begin{figure}[b]
\includegraphics[width=0.95\textwidth]{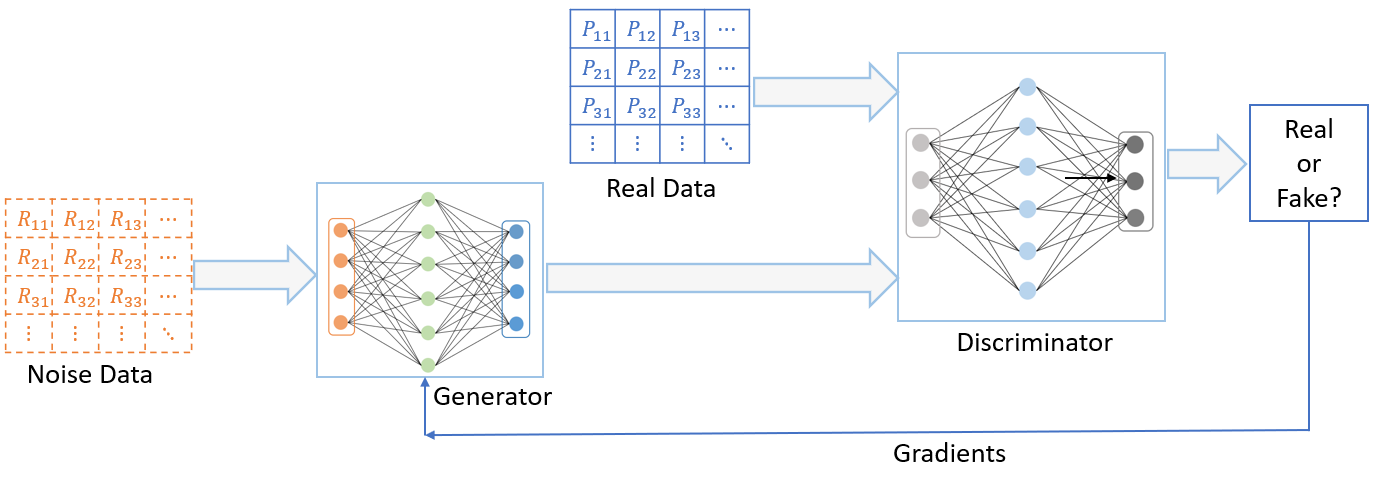}
\caption{Diagram of using generative adversarial network (GAN) model for data simulation.} \label{GAN}
\end{figure}

As illustrated in Figure \ref{GAN}, the architecture of Generative Adversarial Network (GAN) consists of two distinct neural networks: the Generator and the Discriminator. The Generator, often a type of neural network like a convolutional neural network (CNN), is designed to create synthetic data samples. It is denoted as \(\boldsymbol{G}(\cdot)\). The Discriminator, typically another neural network which can also be a CNN (though its structure may vary based on the specific application), is tasked with evaluating whether the data samples are real (authentic data) or fabricated by the Generator. It is denoted as \(\boldsymbol{D}(\cdot)\). In the process (Figure \ref{GAN}), the Generator starts with a noise sample, usually drawn from a Gaussian distribution, which has dimensions compatible with the original data distribution. This noise sample serves as the initial input for the Generator, resulting in \(Simulated\ Data\). Both this \(Simulated\ Data\) and the \(Real\ Data\) are then fed into the Discriminator \(\boldsymbol{D}(\cdot)\). The Discriminator's role is to discern whether each sample is real or simulated. Concurrently, the Generator \(\boldsymbol{G}(\cdot)\) is trained to progressively reduce the difference between the distributions of the real and simulated data through iterative tuning.

\begin{figure}[htbp!]
\centering
\includegraphics[width=0.95\textwidth]{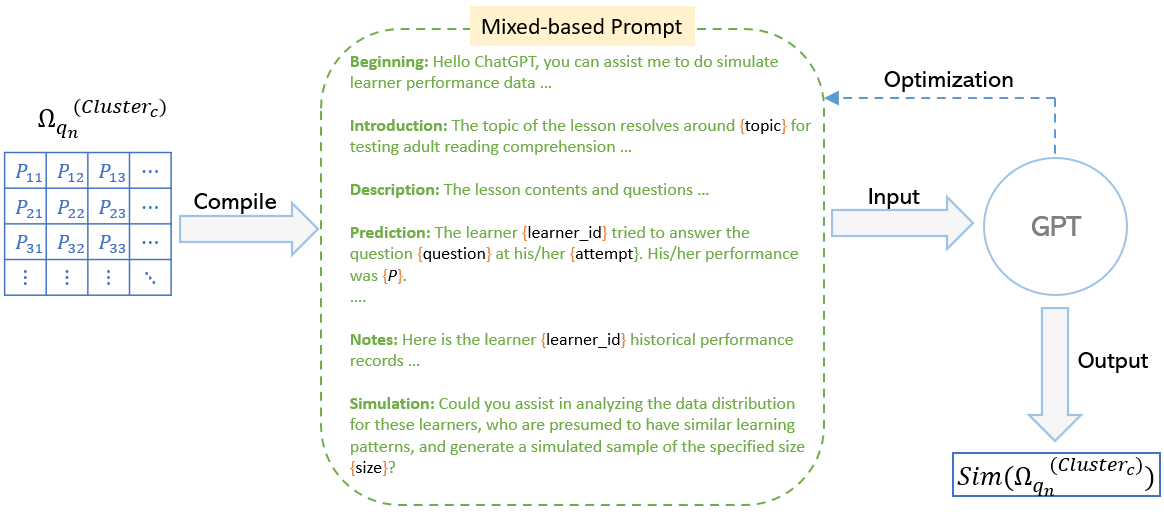}
\caption{Diagram of using generative pre-trained transformer (GPT) model for data simulation.} \label{GPT}
\end{figure}

Considering the limitations of using purely numerical values for interoperability and the enhanced semantic understanding that detailed descriptions provide, we have developed a mixed-based prompt approach for GPT-4 (illustrated in Figure \ref{GPT}). The prompt strategy integrates original matrix data with interpretive text, thereby enriching the context and interpretability of the data. Additionally, it incorporates the Chain-of-Thought (CoT) prompting technique \cite{wei2022chain}, which involves appending guiding phrases such as \textit{'Let's think step by step'} at the end of the prompt to facilitate a more structured analytical process. Specifically, the constructed prompt includes comprehensive elements such as the reading material being analyzed, detailed information about the questions (including their answers), and the learners-attempts matrix data, complete with descriptive information about both its format and entries. Subsequently, a simulation request prompts GPT-4 to integrate the numerical and textual data in a coherent and insightful manner, ultimately driving the execution of a simulation. During the optimization process, these prompts are iteratively refined and adjusted to efficiently yield results that align with our specified objectives.


\section{Results}

\begin{table}[b]
\centering
\caption{Results about the sparsity levels and latent features by tensor completion. \textbf{M}, \textbf{H}, and \textbf{E} denote \textbf{M}edium, \textbf{H}ard, and \textbf{E}asy lesson levels, respectively.}\label{tab1}
    \begin{tabular}{cccccc}
        \toprule
        Dataset & Sparsity Level (Original) & \(K\) (Latent Features)  \\
        \midrule
        Lesson (M) & 84.02\% & 6 \\
        Lesson (H) & 85.45\% & 6 \\
        Lesson (E) & 81.25\% & 4 \\
        \bottomrule
    \end{tabular}
\smallskip \\
\begin{minipage}{\linewidth}
\end{minipage}
\end{table}

As illustrated in Table \ref{tab1}, the original dataset exhibits sparsity levels ranging from 80\% to 85\% (as determined by calculating the proportion of missing values to the total number of entries). By iteratively tuning the latent feature range 
[1, 20] in tensor factorization algorithms, we identified the optimal number of latent features (\(K\)) as 6 for both Lesson (M) and Lesson (H), and 4 for Lesson (E). The optimal \(K\) value was derived by averaging results from multiple trainings with optimized \(K\) values in tensor factorization.

These findings suggest that tensor completion (based on tensor factorization) can efficiently impute missing values in the original sparse performance data, notably for unexplored questions and attempts. This enhancement is crucial for facilitating more comprehensive analysis and modeling in Intelligent Tutoring Systems (ITSs). The latent features, closely associated with learner-specific characteristics during the learning process, are captured with nuanced detail, particularly in the context of reading comprehension. Further research is imperative to fully understand the underlying physical essence of these latent features.

\begin{figure}[b]
    \centering
    \begin{subfigure}[t]{6in}
        \includegraphics[width=1\textwidth]{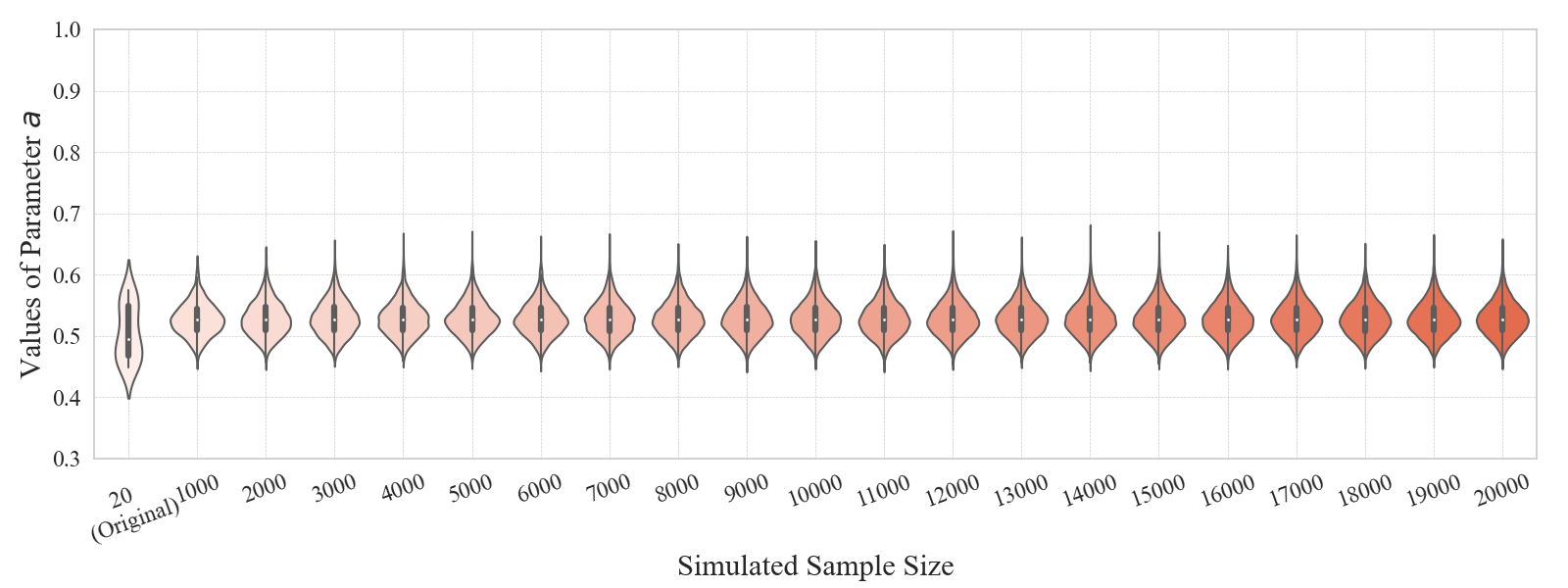}
        \caption{Distribution of parameter \(a\) by GAN simulation.}
        \label{fig:GAN_a}
    \end{subfigure}
    \hfil
    \begin{subfigure}[t]{6in}
        \includegraphics[width=1\textwidth]{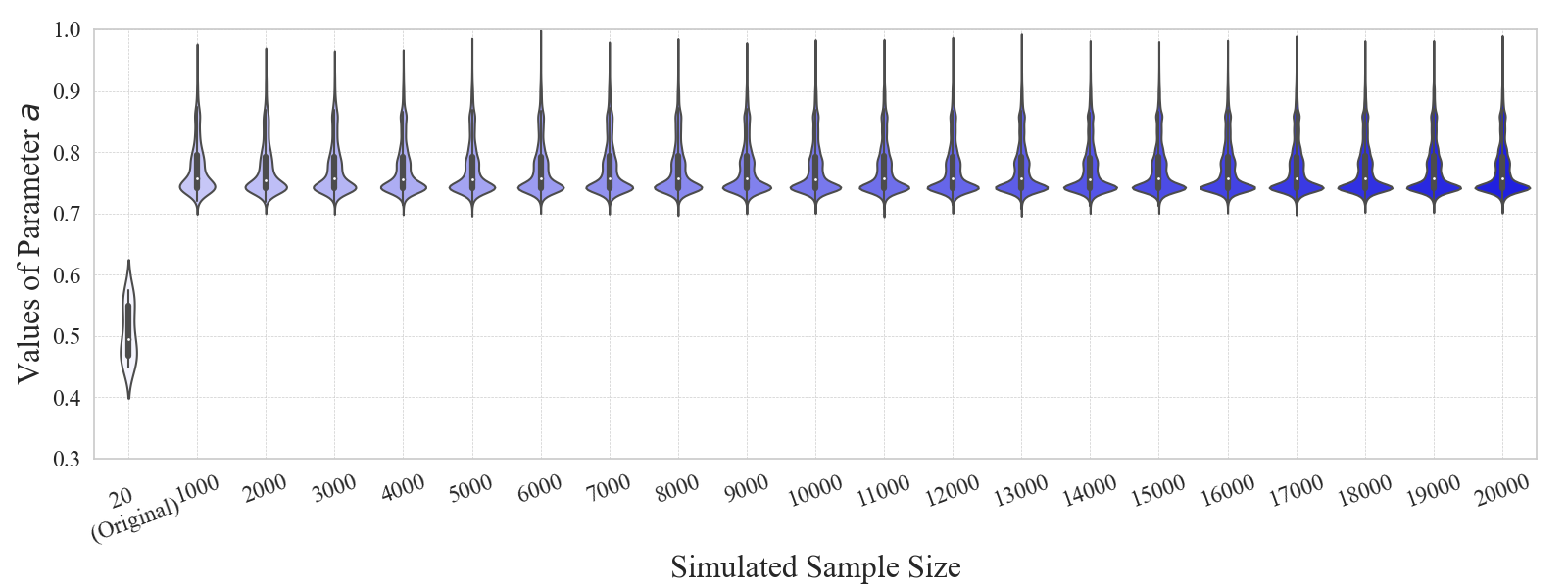}
        \caption{Distribution of parameter \(a\) by GPT-4 simulation.} 
        \label{fig:GPT_a}
    \end{subfigure}
    \caption{Distribution of parameter \(a\) by simulation.} 
    \label{fig:parameter_a_evaluation}
\end{figure} 

\begin{figure}[h]
    \centering
    \begin{subfigure}[t]{6in}
        \includegraphics[width=1\textwidth]{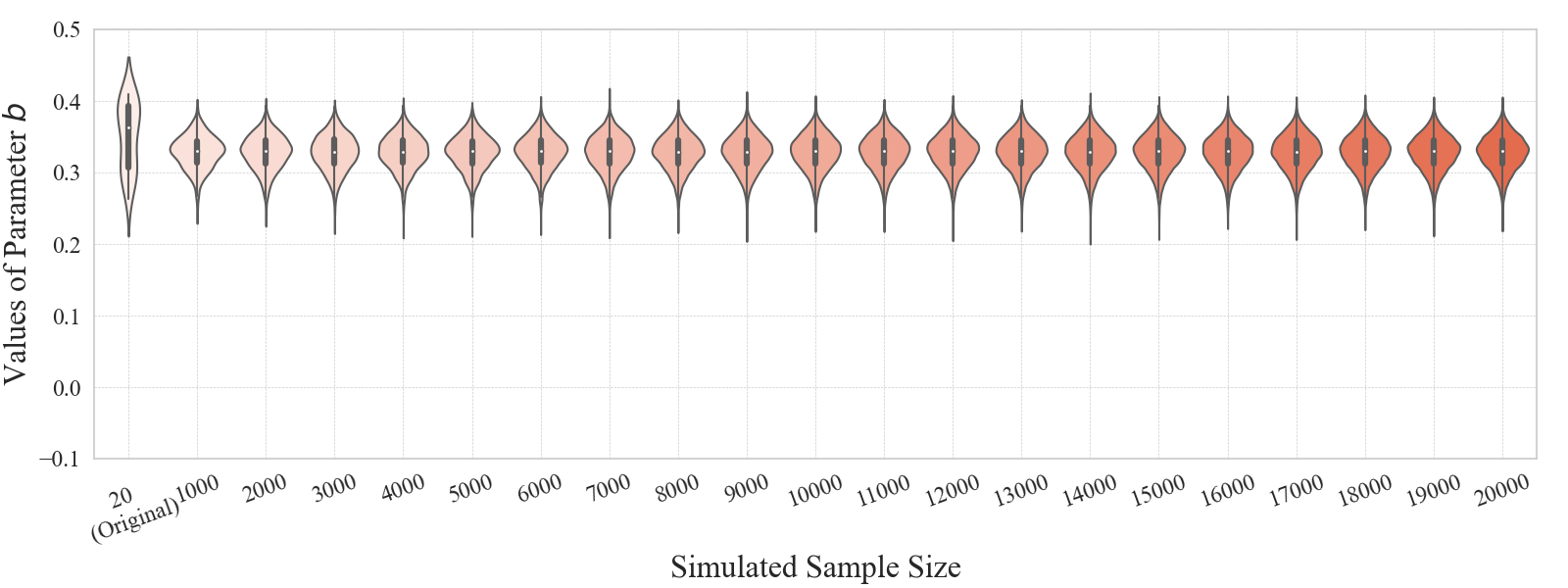}
        \caption{Distribution of parameter \(b\) by GAN simulation.}
        \label{fig:GAN_b}
    \end{subfigure}
    \hfil
    \begin{subfigure}[t]{6in}
        \includegraphics[width=1\textwidth]{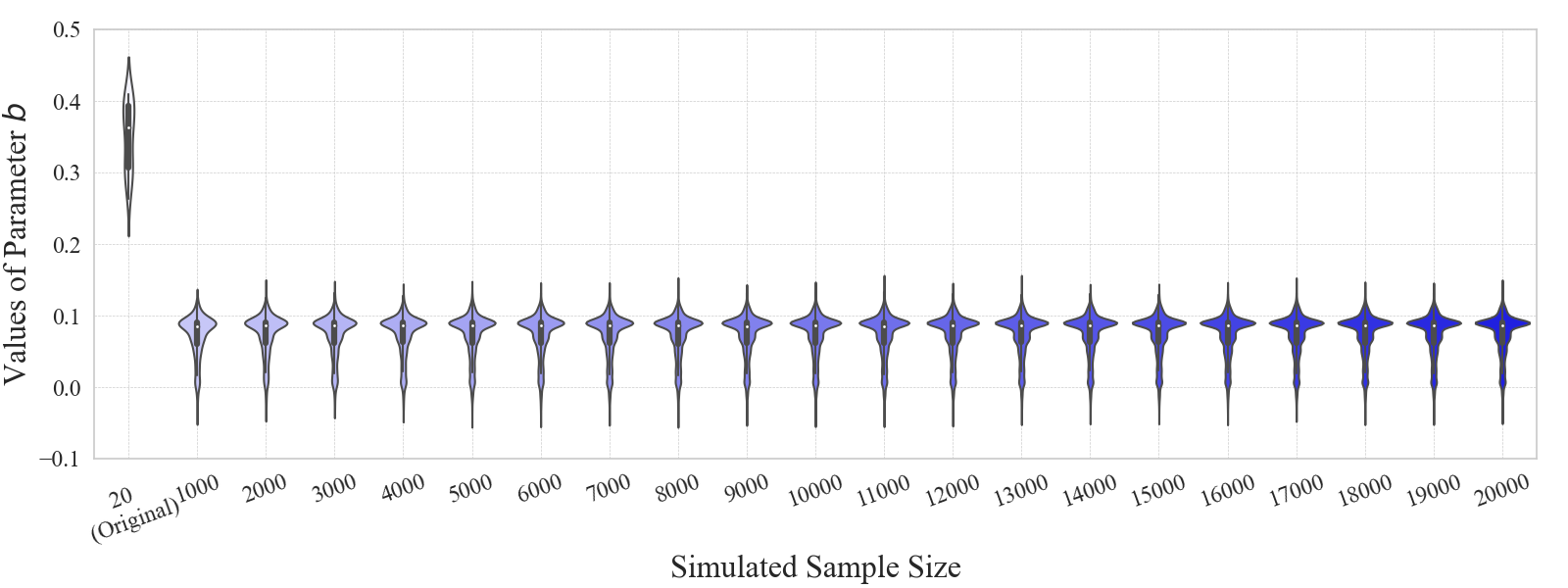}
        \caption{Distribution of parameter \(b\) by GPT-4 simulation.} 
        \label{fig:GPT_b}
    \end{subfigure}
    \caption{Distribution of parameter \(b\) by simulation.} 
    \label{fig:parameter_b_evaluation}
\end{figure}

The distributions of parameters \(a\) and \(b\) are illustrated in Figure \ref{fig:parameter_a_evaluation} and Figure \ref{fig:parameter_b_evaluation}, respectively. These figures visualize the parameter distributions from a example cluster data set with an original size of 20, and exhibit simulations in increments of 1000, with total sizes ranging from 1000 to 20000. 

Figure \ref{fig:parameter_a_evaluation} demonstrates the distributions of parameters \(a\), which is used to represent  the learner’s initial ability or prior knowledge.
Figure \ref{fig:GAN_a} shows the distribution of the parameter \(a\) obtained by GAN simulation. As the sample size increases, the range of parameter \(a\) from the simulation sample mostly falls within the original range of parameter \(a\), although it exhibits a longer tail distribution extending beyond the original maximum value of parameter \(a\). The distribution of the parameter \(a\) obtained by GPT-4 simulation is illustrated in Figure \ref{fig:GPT_a}. Unlike those obtained from GAN simulation, the range of parameter \(a\) values here extends beyond the original range, which is particularly evident as the simulated sample size increases. This suggests that the initial learning ability in GPT-4 simulated samples exhibits more variability and divergence from the original data compared to those from GAN simulation. 

Then, Figure \ref{fig:parameter_b_evaluation} demonstrates the distributions of parameters \(b\) as derived from both GAN and GPT-4 simulations. The parameter \(b\) represents the learning rate, which reflects how quickly a learner acquires knowledge through practice. The GAN simulation produces a narrower range of parameter \(b\) values, especially in terms of maximum and minimum values, when compared to the original dataset, as depicted in Figure \ref{fig:GAN_b}. With increasing sample size, this range generally maintains a consistent pattern. On the other hand, the GPT-4 simulation, as shown in Figure \ref{fig:GPT_b}, demonstrates a broader range for parameter \(b\), extending beyond the original scope. This contrast suggests that GPT-4 simulation may capture a wider variability in learning rates compared to GAN simulation.

\section{Discussion and Conclusion}

This paper proposed the 3DG systematic simulation framework based on generative models (particularly GAN and GPT) to address data sparsity challenges in learning performance data within intelligent tutoring systems (ITS). The framework involves representing learner data from problem-solving step attempts as a three-dimensional tensor with the axes of learners, questions, and attempts. Tensor completion, based on tensor factorization, is then utilized to impute missing performance data entries, generating a dense tensor. Such imputation computation leverages the similarities in learner performance across various questions and attempts, capturing the sequential and temporal dynamics of learning \cite{conway2001sequential,conway2012}. We have demonstrated the integration of generative models, including GAN and GPT-4 for creating scalable, individualized learning simulations aimed at enhancing learner models for personalized instruction. Our comparative analysis reveals that GAN surpasses GPT-4 in terms of reliability for scalable simulations. 

Overall, the GAN simulations demonstrate a narrower and more consistent range of values for parameters \(a\) and \(b\), indicating higher reliability for scalable simulations compared to the broader value range exhibited by the GPT-4 simulations. The mechanism for the GPT-4 simulations, refined through iterative optimization of GPT-4 prompts, involves selecting random values from a flat array of the original data. These values are then adjusted to match base probabilities, preserving the overall data distribution while facilitating the creation of an expanded dataset. Although valuable in computational simulations, this method generally underperforms in numerical computing compared to deep learning models, as demonstrated by the GAN's performance in this study.

Our findings shed light on the potential use of GPT-4 in simulating learner performance represented through numerical values in future research. \textit{Firstly}, employing mixed-based prompts improves interoperability with numerical data, thus enhancing the efficiency of subsequent modeling and simulation computations. \textit{Secondly}, the Chain-of-Thought (CoT) prompting technique delineates the steps for the simulation task, effectively directing GPT-4 in its reasoning process. This includes a structured approach comprising: \textbf{Understanding the Existing Matrix}, \textbf{Distribution Analysis}, \textbf{Clustering Information}, and \textbf{Simulation Process}. \textit{Thirdly}, the computational power of GPT-4 in modeling and simulation is attributed to its capabilities in  self-search, self-programming, and self-computing, all of which are facilitated by prompt engineering. This significantly enhances its utility in data analysis and modeling for future research endeavors. However, integrating GPT-4 with numerical computation presents fundamental challenges, as we discuss in the following section.

\section{Limitations and Future Works}

The capability of GPT-4 in performing deep learning tasks involving numerical computations remains insufficient, primarily due to the intrinsic limitations of large language models and platform contraints. Future research could productively explore the integration of GAN with GPT models, aiming to improve their interoperability and computational capabilities. Furthermore, the degree of sparsity in the original performance data, particularly when formatted as a tensor, significantly impacts the performance of generative models. Therefore, investigating the sensitivity and robustness of tensor completion methods in response to different levels of data sparsity presents an important avenue for future studies. Such investigations are crucial for better integrating large language models within Intelligent Tutoring Systems (ITSs), potentially leading to more refined and effective educational tools.

\section{Acknowledgements}

We extend our sincere gratitude to Prof. Philip I. Pavlik Jr. from the University of Memphis and Prof. Shaghayegh Sahebi from the University at Albany - SUNY for their expert guidance on tensor factorization method. 


\bibliography{sample-ceur}

\end{document}